\documentclass[12pt]{article}
\usepackage{psfrag} 
\usepackage{graphics}
\usepackage{amssymb,epsfig,amsmath,euscript,array,cite}

\newcounter{multieqs}

\newcommand{\be}{\begin{equation}}
\newcommand{\ee}{\end{equation}}
\newcommand{\eq}[1]{(\ref{#1})}

\newcommand{\bra}[1]{\langle #1|}
\newcommand{\ket}[1]{|#1 \rangle}

\newcommand{\bm}[1]{\mbox{\boldmath $#1$}}

\def\bd{\begin{document}}
\def\ed{\end{document}}
\def\nn{\nonumber}
\def\bea{\begin{eqnarray}}
\def\eea{\end{eqnarray}}
\let\bm=\bibitem
\let\la=\label

\def\npb#1#2#3{Nucl. Phys. {\bf{B#1}} #3 (#2)}
\def\plb#1#2#3{Phys. Lett. {\bf{#1B}} #3 (#2)}
\def\prl#1#2#3{Phys. Rev. Lett. {\bf{#1}} #3 (#2)}
\def\prd#1#2#3{Phys. Rev. {D \bf{#1}} #3 (#2)}
\def\cmp#1#2#3{Comm. Math. Phys. {\bf{#1}} #3 (#2)}
\def\cqg#1#2#3{Class. Quantum Grav. {\bf{#1}} #3 (#2)}
\def\nppsa#1#2#3{Nucl. Phys. B (Proc. Suppl.) {\bf{#1A}}#3 (#2)}
\def\ap#1#2#3{Ann. of Phys. {\bf{#1}} #3 (#2)}
\def\ijmp#1#2#3{Int. J. Mod. Phys. {\bf{A#1}} #3 (#2)}
\def\rmp#1#2#3{Rev. Mod. Phys. {\bf{#1}} #3 (#2)}
\def\mpla#1#2#3{Mod. Phys. Lett. {\bf A#1} #3 (#2)}
\def\jhep#1#2#3{J. High Energy Phys. {\bf #1} #3 (#2)}
\def\atmp#1#2#3{Adv. Theor. Math. Phys. {\bf #1} #3 (#2)}

\newcommand{\EQ}[1]{\begin{equation} #1 \end{equation}}
\newcommand{\AL}[1]{\begin{subequations}\begin{align} #1 \end{align}\end{subequations}}
\newcommand{\SP}[1]{\begin{equation}\begin{split} #1 \end{split}\end{equation}}
\newcommand{\ALAT}[2]{\begin{subequations}\begin{alignat}{#1} #2 \end{alignat}\end{subequations}}
\def\beqa{\begin{eqnarray}} 
\def\eeqa{\end{eqnarray}} 
\def\beq{\begin{equation}} 
\def\eeq{\end{equation}} 

\def\N{{\cal N}}
\def\sst{\scriptscriptstyle}
\def\thetabar{\bar\theta}
\def\Tr{{\rm Tr}}
\def\one{\mbox{1 \kern-.59em {\rm l}}}

\def\a{\alpha}		\def\da{{\dot\alpha}}
\def\b{\beta}		\def\db{{\dot\beta}}
\def\c{\gamma} 	\def\C{\Gamma}	\def\cdt{\dot\gamma}
\def\d{\delta}	\def\D{\Delta}	\def\ddt{\dot\delta}
\def\e{\epsilon}		\def\vare{\varepsilon}
\def\f{\phi}	\def\F{\Phi}	\def\vvf{\f}
\def\h{\eta}
\def\k{\kappa}
\def\l{\lambda}	\def\L{\Lambda}
\def\m{\mu}	\def\n{\nu}
\def\o{\omega}
\def\p{\pi}	\def\P{\Pi}
\def\r{\rho}
\def\s{\sigma}	\def\S{\Sigma}
\def\t{\tau}
\def\th{\theta}	\def\Th{\Theta}	\def\vth{\vartheta}
\def\X{\Xeta}
\def\z{\zeta}

\def\cA{{\cal A}} \def\cB{{\cal B}} \def\cC{{\cal C}}
\def\cD{{\cal D}} \def\cE{{\cal E}} \def\cF{{\cal F}}
\def\cG{{\cal G}} \def\cH{{\cal H}} \def\cI{{\cal I}}
\def\cJ{{\cal J}} \def\cK{{\cal K}} \def\cL{{\cal L}}
\def\cM{{\cal M}} \def\cN{{\cal N}} \def\cO{{\cal O}}
\def\cP{{\cal P}} \def\cQ{{\cal Q}} \def\cR{{\cal R}}
\def\cS{{\cal S}} \def\cT{{\cal T}} \def\cU{{\cal U}}
\def\cV{{\cal V}} \def\cW{{\cal W}} \def\cX{{\cal X}}
\def\cY{{\cal Y}} \def\cZ{{\cal Z}}

\def\ua{\underline{\alpha}}
\def\ub{\underline{\phantom{\alpha}}\!\!\!\beta}
\def\uc{\underline{\phantom{\alpha}}\!\!\!\gamma}
\def\um{\underline{\mu}}
\def\ud{\underline\delta}
\def\ue{\underline\epsilon}
\def\una{\underline a}\def\unA{\underline A}
\def\unb{\underline b}\def\unB{\underline B}
\def\unc{\underline c}\def\unC{\underline C}
\def\und{\underline d}\def\unD{\underline D}
\def\une{\underline e}\def\unE{\underline E}
\def\unf{\underline{\phantom{e}}\!\!\!\! f}\def\unF{\underline F}
\def\unm{\underline m}\def\unM{\underline M}
\def\unn{\underline n}\def\unN{\underline N}
\def\unp{\underline{\phantom{a}}\!\!\! p}\def\unP{\underline P}
\def\unq{\underline{\phantom{a}}\!\!\! q}
\def\unQ{\underline{\phantom{A}}\!\!\!\! Q}
\def\unH{\underline{H}}

\def\As {{A \hspace{-6.4pt} \slash}\;}
\def\bs {{b \hspace{-6.4pt} \slash}\;}
\def\Ds {{D \hspace{-6.4pt} \slash}\;}
\def\ds {{\del \hspace{-6.4pt} \slash}\;}
\def\ss {{\s \hspace{-6.4pt} \slash}\;}
\def\ks {{ k \hspace{-6.4pt} \slash}\;}
\def\ps {{p \hspace{-6.4pt} \slash}\;}
\def\pas {{{p_1} \hspace{-6.4pt} \slash}\;}
\def\pbs {{{p_2} \hspace{-6.4pt} \slash}\;}

\def\Fh{\hat{F}}
\def\Vh{\hat{V}}
\def\Xh{\hat{X}}
\def\ah{\hat{a}}
\def\xh{\hat{x}}
\def\yh{\hat{y}}
\def\ph{\hat{p}}
\def\xih{\hat{\xi}}

\def\psit{\tilde{\psi}}
\def\Psit{\tilde{\Psi}}
\def\tht{\tilde{\th}}
 
\def\At{\tilde{A}}
\def\Qt{\tilde{Q}}
\def\Rt{\tilde{R}}
\def\Nt{\tilde{N}}

\def\at{\tilde{a}}
\def\st{\tilde{s}}
\def\ft{\tilde{f}}
\def\pt{\tilde{p}}
\def\qt{\tilde{q}}
\def\vt{\tilde{v}}
\def\nt{\tilde{n}}

\def\delb{\bar{\partial}}
\def\bz{\bar{z}}
\def\bD{\bar{D}}
\def\bB{\bar{B}}

\def\bk{{\bf k}}
\def\bl{{\bf l}}
\def\bp{{\bf p}}
\def\bq{{\bf q}}
\def\br{{\bf r}}
\def\bx{{\bf x}}
\def\by{{\bf y}}
\def\bR{{\bf R}}
\def\bV{{\bf V}}

\def\d{\delta}\def\D{\Delta}\def\ddt{\dot\delta}

\def\pa{\partial} \def\del{\partial}
\def\xx{\times}
\def\uno{\mbox{1 \kern-.59em {\rm l}}}  

\def\trp{^{\top}}
\def\inv{^{-1}}
\def\dag{{^{\dagger}}}
\def\pr{^{\prime}}

\def\rar{\rightarrow}
\def\lar{\leftarrow}
\def\lrar{\leftrightarrow}

\newcommand{\0}{\,\!}      
\def\one{1\!\!1\,\,}
\def\im{\imath}
\def\jm{\jmath}

\newcommand{\tr}{\mbox{tr}}
\newcommand{\slsh}[1]{/ \!\!\!\! #1}

\def\vac{|0\rangle}
\def\lvac{\langle 0|}

\def\hlf{\frac{1}{2}}
\def\ove#1{\frac{1}{#1}}

\def\Box{\square}
\def\ZZ{\mathbb{Z}}
\def\CC#1{({\bf #1})}
\def\bcomment#1{}
\def\bfhat#1{{\bf \hat{#1}}}
\def\VEV#1{\left\langle #1\right\rangle}

\newcommand{\ex}[1]{{\rm e}^{#1}} \def\ii{{\rm i}}

\begin{document}

\hfill{hep-th/0206005}
 
\vspace{20pt}
 
\begin{center}

{\Large \bf Three-point functions in $\cN=4$ Yang-Mills theory }

{\Large \bf and pp-waves} \vspace{30pt}

\vspace{30pt}
 
{\bf Chong-Sun Chu, Valentin V. Khoze
and Gabriele Travaglini}

{\small \em Centre for Particle Theory, 
University of Durham,  \\Durham, DH1 3LE, UK}

Email: {\sffamily \tt chong-sun.chu, valya.khoze,
gabriele.travaglini@durham.ac.uk }

\vspace{30pt}
{\bf Abstract}

\end{center}
Recently it has been proposed that the coefficient of the three-point
function of the BMN operators in $\cN=4$ supersymmetric Yang-Mills
theory is related to the three-string interactions in the pp-wave background.
We calculate three-point functions of these operators to the first
order in the effective Yang-Mills coupling $\l' = g_{\rm YM}^2 N/J^2$
in planar perturbation theory. On the string theory side, we derive
the explicit 
expressions of the Neumann matrices
to all orders in $1/(\mu p^+ \a')^2$. This allows us to 
compute the corresponding three-string scattering amplitudes.
This provides an all orders prediction for the field
theory three-point functions. 
We compare our field theory results 
with the string theory results to the subleading order in
$1/(\mu p^+ \a')^2$ and find perfect agreement. 

\vspace{0.5cm}

\setcounter{page}0
\thispagestyle{empty}
\newpage


\section{Introduction}

Berenstein, Maldacena and Nastase (BMN)
put forward an insightful proposal towards understanding of
the massive string modes \cite{BMN}. The field theory side of this 
correspondence is the conformal $\cN=4$ SYM in a new double scaling
limit. The nature of this double scaling limit was further elucidated
in \cite{seme,dzf,gross}.

Crucial to the BMN line of reasoning is 
the emergence of the pp-wave background,
which arises as a Penrose limit \cite{penrose} of  $AdS_5\times S^5$
\cite{bfhp1,bfhp2,bfhp3}. This background is maximally supersymmetric and
remarkably string theory in this background is exactly solvable, see
\cite{s1,s2} for closed strings, and \cite{op1,op2,op3} for the open string case. 
First attempts 
towards understanding of the 
holographic relation in the pp-wave context were made in \cite{hr1,hr2,hr3,hr4}.
The pp-wave/SYM correspondence of BMN goes significantly beyond the original
AdS/CFT duality \cite{ads1,ads2,ads3,rev1,rev2}  
since it provides a map between field theory operators and
generic string states in the lightcone gauge, 
and not just the supergravity multiplet
($n=0$).  For example, 
\begin{equation} \label{bmn1}
\frac{1}{\sqrt{J} N^{J/2+1}} \Tr Z^J  \longleftrightarrow \ket{0,p^+},
\end{equation}
\begin{equation}\label{bmn2}
\frac{1}{\sqrt{J} N^{J/2+1}} \sum_{l=0}^J \Tr [\phi^3 Z^l \phi^4
Z^{J-l}] e^{\frac{2 \pi i n l}{J} } \longleftrightarrow a^{7 \dag}_n
a^{8 \dag}_{-n} \ket{0,p^+}  .
\end{equation}
These relations give the string vacuum and the first excited string
states in terms of SYM operators.  Here $N$ is the number of colours and
$Z= \phi_5+ i\phi_6$ is a complex scalar field with
unit $R$ charge. The scaling dimensions $\D$ of the  Yang-Mills 
operators in \eq{bmn1}, \eq{bmn2} are related to the masses of the
corresponding string states via
\begin{equation}
\Delta -J= H_{\rm lc}/\mu,
\end{equation}
where $H_{\rm lc}$ is the lightcone string Hamiltonian and $\mu$ is the
scale of the pp-wave metric. Written in terms of gauge theory
parameters, this gives a prediction for the conformal dimension of
the BMN operators
\begin{equation} \label{dJ}
\Delta -J = \sqrt{1+ \frac{g_{\rm YM}^2 N n^2}{J}}.
\end{equation}

The BMN correspondence is understood to hold in 
the double scaling limit: 
 \begin{equation} \label{double} 
N \to \infty\ , \quad J \sim \sqrt{N} \quad
\,\mbox{with\, $g_{\rm YM}$\, fixed}. 
\end{equation} 
In this limit the 't Hooft
coupling $\lambda=g_{\rm YM}^2 N$ is infinite and perturbative
calculations in gauge theory are hopeless in general. A well known
exception to this rule concerns
BPS operators which receive no perturbative
corrections at all, and their scaling dimensions $\Delta$ coincide
with their free theory engineering dimensions $\Delta^0$.  
BMN
instead considered a class of `near-BPS' operators, as in \eq{bmn2} 
with large $R$-charge $J$, which do receive quantum corrections, but these
operators are constructed in such a way that in perturbative
evaluation of their scaling dimensions, $\lambda$ is accompanied
by a suppression factor $1/J^2$. For these operators the coupling
is effectively  
\begin{equation} \label{lampr}
 \l' = \frac{g_{\rm YM}^2 N}{J^2} = \frac{1}{(\mu p^+ \a')^2}\ , 
\end{equation} 
which is  finite in the large $N$ limit \eq{double} and can be taken small
\cite{BMN}, see also \cite{seme,dzf,gross}. 

It was further assumed
in \cite{BMN} that the gauge theory remains planar in the limit
\eqref{double} for the class of BMN operators even though the
original 't Hooft coupling $\lambda$ is infinite. Non-planar
diagrams in the BMN limit \eq{double} were first studied in \cite{seme} and in
\cite{dzf} and were found to be important and governed by
$J^4/N^2$. It follows from the double scaling limit \eqref{double}
that in addition to $\l'$ defined in \eqref{lampr}, there is a
second dimensionless constant
\begin{equation} 
g_2 :=\frac{J^2}{N}= 4 \pi g_s (\mu p^+ \a')^2 \ , 
\end{equation} 
which plays the r{\^o}le of the genus counting parameter as explained in
\cite{seme,dzf}.

In this paper, we will be mostly concerned with the following BMN operators:
\begin{equation}\label{o3}
\cO^J_{\rm vac} := \frac{1}{\sqrt{J N^{J}}} \tr (Z^J),
\end{equation} 
\begin{equation}\label{o2}
\cO^J_0 := \frac{1}{\sqrt{N^{J+1}}} \tr (\Phi Z^J),
\end{equation}
\begin{equation} \label{o1}
\cO^J_{n,-n} := \frac{1}{\sqrt{J N^{J+2}}} \sum_{l=0}^J \tr(\Phi Z^l \Psi
Z^{J-l}) e^{\frac{2 \pi i n l}{J}}
\end{equation}
and their two and three-point correlation functions. Here
\begin{equation}
\Phi = \phi_1+ i \phi_2, \quad \Psi = \phi_3+ i \phi_4,
\quad Z = \phi_5+ i \phi_6
\end{equation}
are the three complex scalar fields of the $\cN=4$ theory.
In perturbation theory, the flavour of $\Phi$, $\Psi$ and $Z$ is conserved.  
The operators $\cO^J_{\rm vac}, \cO^J_0 ,\cO^J_{0,0}$ 
are half BPS and correspond to the string vacuum and supergravity
states. The operator $\cO^J_{n,-n}$ for $n \neq 0$ is non-BPS and
receives quantum corrections to its scaling dimension.
At the level of planar diagrams, BMN operators do not mix 
\cite{seme,dzf,gross}
and have well-defined conformal dimensions.
Two and three-point functions of chiral operators with arbitrary R-charges
had been calculated in \cite{ram} and \cite{seme}. 
It follows from the conformal
invariance  of the theory that the two-point function can be written 
in the canonical  form 
\footnote{The operators in \eq{o3}, \eq{o2}, \eq{o1} are already
normalized such that \eq{2pt} holds in free field theory.}
\begin{equation} \label{2pt}
\langle {\cO}_i (0) \cO_j(x) \rangle = \frac{\d_{ij}}{(4 \pi^2
x^2)^{\Delta_i}}. 
\end{equation}
 Furthermore conformal invariance
implies that the three-point function  takes the form
\begin{equation} \label{3pt}
\langle \cO_{i_1}(x_1) \cO_{i_2}(x_2) \cO_{i_3}(x_3) \rangle  =
\frac{C_{{i_1} {i_2} {i_3}} }
{(4\pi^2 x_{12}^2)^{\frac{\D_1+\D_2 -\D_3}{2}}
 (4\pi^2 x_{13}^2)^{\frac{\D_1+\D_3 -\D_2}{2}}
 (4\pi^2 x_{23}^2)^{\frac{\D_2+\D_3 -\D_1}{2}}
}
\end{equation}
where $x_{ij}^2: = (x_i-x_j)^2$.
When nonplanar diagrams are taken into account, BMN operators
$\cO^J_{n,-n}$ with different nonvanishing values of $n$ mix with each
other already in free field theory \cite{seme,dzf}. Hence the original
BMN operators do not have well defined conformal dimensions and one
has to define a new basis of such operators which does not mix \cite{seme}.
This redefinition has to be implemented order by order in $g_2$ and
$\l'$.  Equations \eq{2pt} and \eq{3pt} represent the correlation
functions of these redefined operators with well-defined
conformal dimensions $\D_i$. The authors of \cite{gross} calculated
anomalous dimensions to the order ${\l'}^2$ at the planar level
(leading order in $g_2$). 
Alternatively, one may also go beyond the original BMN perturbative
computation of anomalous dimensions by including higher genus
diagrams \cite{seme,dzf}. 
Planar three-point functions involving nonchiral operators in free
field theory were calculated in \cite{dzf}. 
In this paper, we  will 
consider the planar limit in order to  work with the original BMN basis of
operators, thus avoiding the complications from operator mixing. 

Due to conformal invariance, all the nontrivial information of the
three-point function is contained in the $x$-independent 
coefficient $C_{i_1 i_2 i_3}$. It is natural to expect that $C_{i_1
i_2 i_3}$ is related to 3-strings interaction in pp-wave
background. One such proposal was put forward in \cite{dzf} and
further analyzed in \cite{lee}. The proposal of \cite{dzf} states that
the matrix element of the lightcone Hamiltonian is related to the
coefficient of the three-point function in field theory via 
\begin{equation}
\bra{i} P^- \ket{j,k} = \mu (\D_i -\D_j -\D_k) C_{ijk} 
\end{equation}
in the leading order in $\l'$. 
Another proposal considered in \cite{huang} relates the ratio of the
three-string amplitudes with those of the field theory three-point
function coefficients 
\begin{equation} \label{hh}
\frac{ \langle \Phi_1| \langle \Phi_2|\langle \Phi_3| V \rangle }
{\langle 0_1| \langle 0_2|\langle 0_3| V\rangle }=
\frac{ C_{123} }{ C^{(\rm vac)}_{123} }.
\end{equation} 
Here $\langle \Phi_1| \langle \Phi_2|\langle \Phi_3| V \rangle $ is
the three-string scattering amplitude in the string field theory
formalism, $ \langle 0_1| \langle 0_2|\langle 0_3| V\rangle$ is the vacuum
amplitude and  $V$ is the lightcone three-string vertex
\cite{sv}. $C_{123}$ (resp. $C^{(\rm vac)}_{123}$ ) is
the three-point function coefficient of the
corresponding BMN operators (resp. of the ``vacuum'' operators
\eq{o3}).
 
All the tests in \cite{dzf,lee,huang} 
were restricted to the free field value of $C_{123}$.
In section 3, we will calculate the three-point
functions in gauge theory 
to the first nontrivial order in $\l'$. 
According to \eq{lampr}, our results correspond to 
string theory in the subleading order in $\l'=1/(\mu p^+ \a')^2$. 
In section 4, we will derive the corresponding three-string amplitudes 
and find perfect agreement with the field theory results 
to  $O(\l')$. The  string computation can be easily generalized
to all orders in $\l'$ and we obtain the exact  form of the Neumann
matrices explicitly. Our string theory results
give an exact  prediction for the field
theory three-point functions.

\section{Two-point function: \\normalization of BMN operators at order $\l'$}

At the first order in $\l'$, the normalization of $\cO^J_{n,-n}$ in
\eq{o1} has to be modified. To determine this normalization, we have
to know precisely the two-point function to order $\l'$.  
The correlation functions of composite operators require UV
regularization. We will 
use  dimensional reduction to $D= 4- 2 \e$ dimensions
and work with Feynman rules in coordinate space. Since we work with
planar diagrams, the group indices are trivial and the 
scalar propagator is given by
\begin{equation}
\D(x) = \frac{\C(1-\e)}{4 \pi^{2-\e} (x^2)^{1-\e}}.
\end{equation}
The scalar four-point interaction can be conveniently divided into
F-terms and D-terms:
\begin{equation} \label{LF}
\cL_F = -2 g_{\rm YM}^2 \tr \left([Z,\Phi][\bar{\Phi},\bar{Z}] +
[Z,\Psi][\bar{\Psi},\bar{Z}] +[\Phi,\Psi][\bar{\Psi},\bar{\Phi}]
\right) , 
\end{equation}
\begin{equation} \label{LD}
\cL_D = - g_{\rm YM}^2 \tr \left( [Z,\bar{Z}]^2 + 
2 [\Phi,\bar{\Phi}][Z,\bar{Z}] +
2 [\Psi,\bar{\Psi}][Z,\bar{Z}] +
2 [\Phi,\bar{\Phi}][\Psi,\bar{\Psi}]
\right) .
\end{equation}
At the planar level and to the first order in $\l'$, 
by inspecting  individual diagrams it is easy to see that
only the F-term interactions \eq{LF} will contribute to the two
and three-point functions; the D-term interactions, as well as the
scalar self energy corrections and gluon exchanges between two
scalars will have a vanishing net contribution. The cancellation in
fact takes place to all orders in the genus expansion as was 
explained in \cite{old,dzf}.

Taking into account one insertion of the vertices from $\cL_F$, and
combining with the free result,  one obtains the  two-point function
in the large $J$ limit
\begin{equation}
G_2(x):= \langle \bar{\cO}^J_{n,-n} (0) \cO^J_{n,-n}(x) \rangle =
\D(x)^{J+2}(1- 8 \pi^2 \d \cdot I(x) )
\end{equation}
where
\begin{equation}
\d= n^2 \l'
\end{equation}
is the anomalous dimension of the operator $\cO^J_{n,-n}$,
and $I(x)$ is  the interaction integral 
with $\D(x)^2$ removed: 
\bea \label{Ix}
I(x) &:=& \left(\frac{\C(1-\e)}{4 \pi^{2-\e}}\right)^2 (x^2)^{2-2\e}
\int \frac{d^{4-2\e}y}{(y^2)^{2-2\e} (y-x)^{2(2-2\e)}}\nn\\
&=& \frac{1}{8\pi^2} (\frac{1}{\e} + \c +1 + \log \pi + \log x^2 + O(\e)).
\eea
We use a subtraction scheme which subtract the $1/\e$ pole together
with a finite part $s$
\begin{equation}  \label{sub}
\frac{1}{\e} + s.
\end{equation}
To comply with  the canonical form \eq{2pt}, 
the properly normalized $\cO^J_{n,-n}$ is given by
\begin{equation}
\cO^J_{n,-n} =  \frac{1+ \frac{\d}{2} (\c +1 -
\log 4 \pi -s)}{\sqrt{J N^{J+2}}} \sum_{l=0}^J \tr(\Phi Z^l \Psi
Z^{J-l}) e^{\frac{2 \pi i n l}{J}} .
\end{equation}
The $\log x$ term on the right hand side of \eq{Ix} was originally 
calculated in \cite{BMN} and this was sufficient to extract 
the anomalous dimension.  Here we have determined the scheme dependent
finite part. 
For physical quantities, the scheme dependence must disappear.
After we normalize the operators according to \eq{2pt}, all the
correlation functions must be scheme independent. In the next section,   
we will calculate $C_{i_1 i_2 i_3} $.

\section{Three-point function}
 
In this section, we will calculate two simple examples of 
three-point functions
\begin{equation}
G_3(x_1,x_2) := \langle \bar{\cO}^J_{n,-n}(0) \, \cO^{J_1}_{0,0}(x_1) \,
\cO^{J_2}_{\rm vac}(x_2) \rangle ,
\end{equation}
\begin{equation}
\tilde{G}_3(x_1,x_2) := \langle \bar{\cO}^J_{n,-n}(0) \, \cO^{J_1}_{0}(x_1) \,
\cO^{J_2}_{0}(x_2) \rangle ,
\end{equation}
to the first order in $\l'$ and at the planar level. Here $J= J_1+ J_2$.

\begin{figure}[ht]
\psfrag{phi}{$\bar{\Phi}$}
\psfrag{psi}{$\bar{\Psi}$}
\psfrag{k}{$k$}
\psfrag{l}{$l$}
\psfrag{x1}{$x_1$}
\psfrag{x2}{$x_2$}
\begin{center}
{\scalebox{1}{
\includegraphics{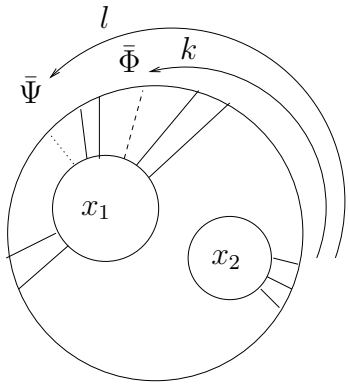}}
}
\end{center} 
\caption{Free diagrams for $G_3$ contributing to $P_1$. The labels $k$
and $l$ count the $Z$-lines as indicated (for the diagram drawn above, 
$k=2$, $l=4$).  }
\label{fig1}
\end{figure}

\begin{figure}[ht]
\label{fig2}
\psfrag{phi}{$\bar{\Phi}$}
\psfrag{psi}{$\bar{\Psi}$}
\psfrag{k}{$k$}
\psfrag{l}{$l$}
\psfrag{x1}{$x_1$}
\psfrag{x2}{$x_2$}
\psfrag{a}{$2a$}
\psfrag{b}{$2b$}
\psfrag{c}{$2c$}
\psfrag{d}{$2d$}
\begin{center}
{\scalebox{1}{
\includegraphics{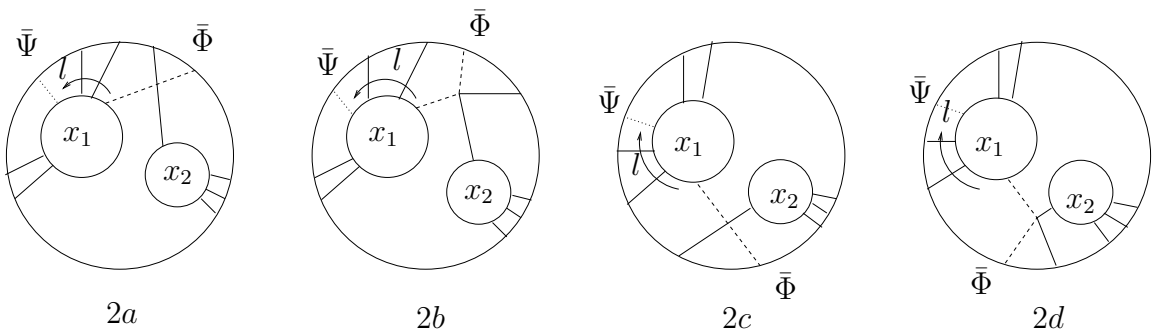}}
}
\end{center} 
\caption{Interacting diagrams for $G_3$ contributing to
$P_2$. Diagrams 2a and 2c have positive signs. Diagrams 2b and 2d have
negative signs.
}
\end{figure}

We first consider $G_3$. The free field theory diagrams are shown in
figure 1 and the surviving interacting diagrams which arise from
the F-term interactions are shown in figure 2. Our result is
given by
\begin{equation} \label{G3}
G_3 =
\frac{J_2}{N \sqrt{J J_1 J_2}} \; [1+ \frac{\d}{2} (\c +1 - \log 4 \pi -s) ]\;
\cdot \D(x_1)^{J_1+2}  \D(x_2)^{J_2}  (P_1 + \l K(x_1,x_2) P_2) .
\end{equation}
The  factor $[1+ \frac{\d}{2} (\c +1 - \log 4 \pi -s)]/(N^{J+2} \sqrt{J J_1
J_2})$ arises from the normalizations of the operators; in addition
summing over the loops of the planar diagram gives rise to a factor of
$N^{J+1}$. Another factor of $J_2$ arises from inequivalent Wick
contractions of $\cO_{\rm vac}$ with the rest. 

Each of the diagrams carries a phase factor which
arises from the operator $\cO_{n,-n}$ and depends on the relative
position of $\Phi$ and $\Psi$. To obtain the total contribution,
one has to sum over all possible positions of  $\Phi$ and $\Psi$ which
amounts to inequivalent ``electrostatic'' diagrams in figures 1, 2, 3 and 4. 
In \eq{G3}, 
$P_1$ and $P_2$ represent the total contributions of these phase
factors  for the free and interacting diagrams of figures 1 and 2
respectively.
The contributions of diagrams  in figures 3 and 4 will be shown
to sum to zero. 
For $P_1$, we have
\begin{equation}
P_1 =  
\sum_{0\leq k,l\leq J_1} e^{\frac{2 \pi i n}{J} (l-k)}
\end{equation}
and  in the large $J$ limit
\begin{equation}
P_1 
=   \frac{J^2}{n^2 \pi^2} \sin^2(\frac{n \pi  J_1}{J}).
\end{equation}
The phase factor
$P_2$ arises from the diagrams with one interaction vertex inserted 
(figure 2). We only need to take into account
interaction vertices arising  from the F-terms  since as mentioned earlier, 
D-term scalar interactions, self energy corrections and gluon
exchanges sum to zero. We find
\bea
P_2 &=& 2(e^{-\frac{2\pi i n}{J}} -1) 
\sum_{l=0}^{J_1} e^{-\frac{2\pi i n l}{J}}
+ 2(e^{\frac{2\pi i n}{J}} -1) 
\sum_{l=0}^{J_1} e^{\frac{2\pi i n l}{J}} \label{P11}\\
&=&  \frac{J^2}{n^2 \pi^2} \sin^2(\frac{n \pi  J_1}{J})
\cdot \frac{-8 \pi^2 n^2}{J^2} . 
\eea
The first term on the right hand side of \eq{P11} comes from the
diagrams 2a and 2b. The relative sign is easily seen from the commutators in 
$\cL_F$. The second term in \eq{P11} comes from the diagrams 2c and
2d, where the $\Phi$
interaction is now at the bottom. The multiplicative factors of 2 in
\eq{P11} arise from summing the diagrams in figure 2 
with $\Phi$ and $\Psi$ exchanged. 

The remaining F-terms diagram are shown in figures 3 and 4 and it is
easy to see that there is a precise cancellation diagram by diagram between
figure 3 and figure 4. For example,  diagram 3a cancels  diagram
4a as they have the same phase factor but  opposite sign.  
Hence these classes of diagrams do not
contribute to $G_3$.

\begin{figure}[ht]
\label{fig3}
\psfrag{phi}{$\bar{\Phi}$}
\psfrag{psi}{$\bar{\Psi}$}
\psfrag{k}{$k$}
\psfrag{l}{$l$}
\psfrag{x1}{$x_1$}
\psfrag{x2}{$x_2$}
\psfrag{a}{$3a$}
\psfrag{b}{$3b$}
\psfrag{c}{$3c$}
\psfrag{d}{$3d$}
\psfrag{e}{$3e$}
\psfrag{f}{$3f$}
\begin{center}
{\scalebox{1}{
\includegraphics{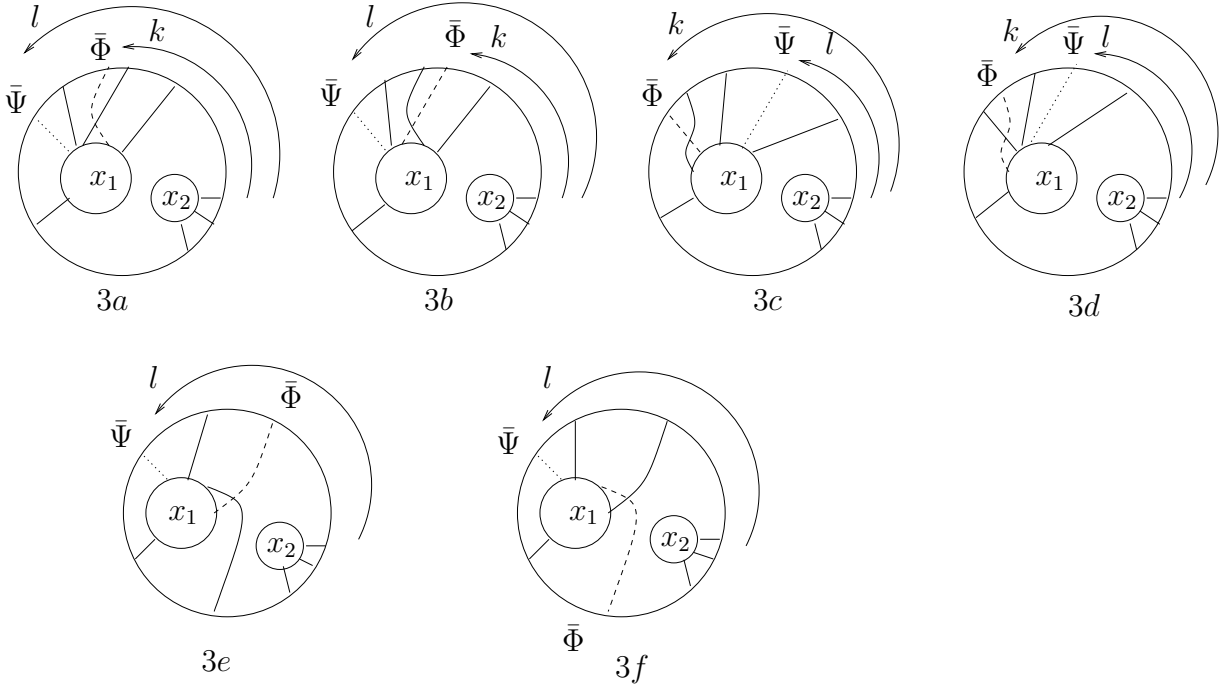}}
}
\end{center} 
\caption{Interacting diagrams for $G_3$. All diagrams come with a
positive sign. }
\end{figure}

\begin{figure}[ht]
\label{fig4}
\psfrag{phi}{$\bar{\Phi}$}
\psfrag{psi}{$\bar{\Psi}$}
\psfrag{k}{$k$}
\psfrag{l}{$l$}
\psfrag{x1}{$x_1$}
\psfrag{x2}{$x_2$}
\psfrag{a}{$4a$}
\psfrag{b}{$4b$}
\psfrag{c}{$4c$}
\psfrag{d}{$4d$}
\psfrag{e}{$4e$}
\psfrag{f}{$4f$}
\begin{center}
{\scalebox{1}{
\includegraphics{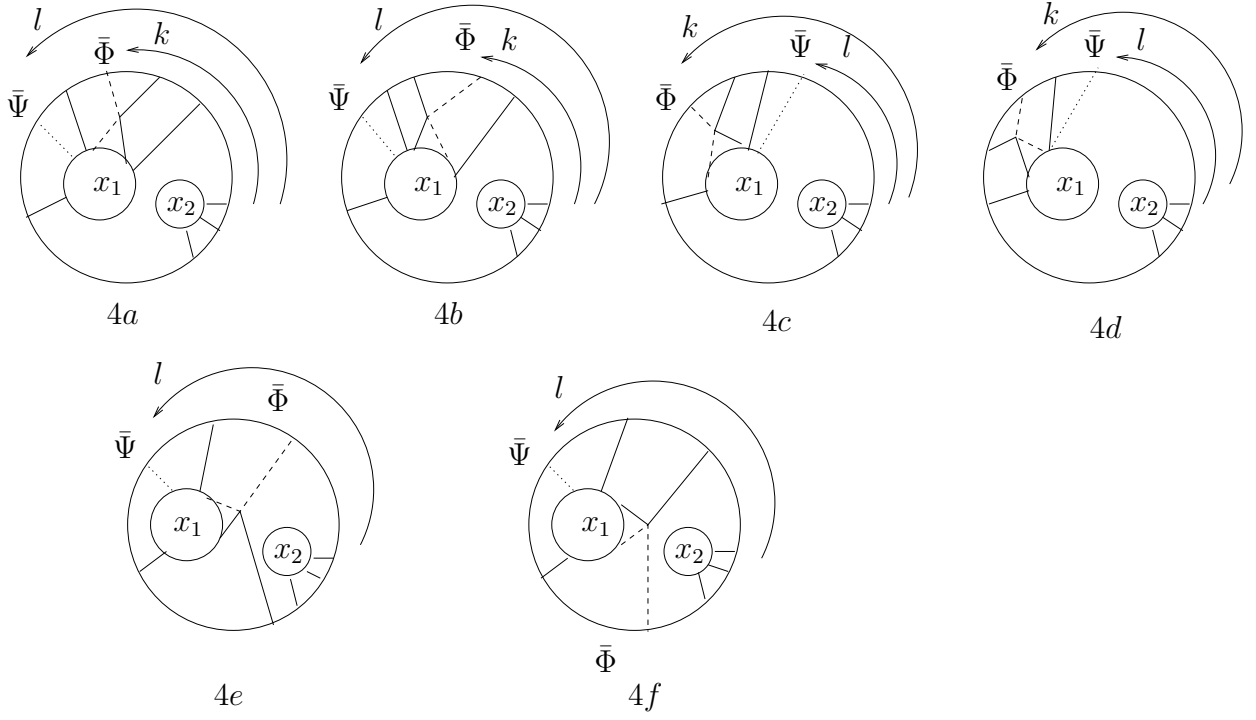}}
}
\end{center} 
\caption{Interacting diagrams for $G_3$. All diagrams come with a
negative sign and precisely cancel those in figure 3.}
\end{figure}

Finally $K(x_1,x_2)$ is the interaction integral for diagrams 2 with
$\D(x_1) \D(x_2)$ removed: 
\begin{equation}
K(x_1,x_2) = \left(\frac{\C(1-\e)}{4 \pi^{2-\e}}\right)^2 
(x_1^2)^{1-\e} (x_2^2)^{1-\e}
\int \frac{d^{4-2\e}y}{(y^2)^{2-2\e} (y-x_1)^{2(1-\e)} 
(y-x_2)^{2(1-\e)}
}
\end{equation}
Evaluating this integral, we obtain
\begin{equation}
K(x_1,x_2) = \frac{1}{16\pi^2} (\frac{1}{\e} + \c +2 + \log \pi + 
\log \frac{x_1^2 x_2^2}{x_{12}^2} + O(\e)).
\end{equation}
Using our subtraction \eq{sub}, we obtain
\begin{equation}
G_3(x_1,x_2) =
\frac{C_{123} }
{
(4\pi^2 x_1^2)^{J_1+ 2 + \frac{\d}{2}}
(4\pi^2 x_2^2)^{J_2+\frac{\d}{2}}
(4\pi^2 x_{12}^2)^{-\frac{\d}{2}}
}
\end{equation}
where the three-point function coefficient is
\begin{equation}
C_{123} = 
\frac{J^{3/2} J_2^{1/2}}{N J_1^{1/2}} 
\frac{1} {n^2 \pi^2} \sin^2(\frac{n \pi J_1}{J}) \; (1- \frac{\l'n^2}{2} ).
\end{equation}
This is one of the main result of this paper and, as anticipated, it
is scheme independent.

\begin{figure}[ht]
\label{fig5}
\psfrag{phi}{$\bar{\Phi}$}
\psfrag{psi}{$\bar{\Psi}$}
\psfrag{k}{$k$}
\psfrag{l}{$l$}
\psfrag{x1}{$x_1$}
\psfrag{x2}{$x_2$}
\begin{center}
{\scalebox{1}{
\includegraphics{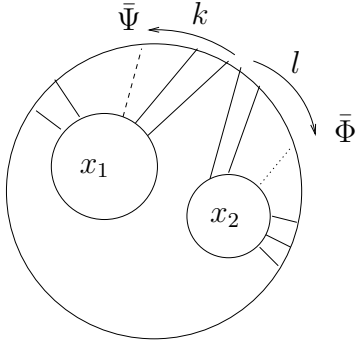}}
}
\end{center} 
\caption{Free diagrams for $\tilde{G}_3$ contributing to $P_5$}
\end{figure}

\begin{figure}[ht]
\label{fig6}
\psfrag{phi}{$\bar{\Phi}$}
\psfrag{psi}{$\bar{\Psi}$}
\psfrag{k}{$k$}
\psfrag{l}{$l$}
\psfrag{x1}{$x_1$}
\psfrag{x2}{$x_2$}
\psfrag{a}{$6a$}
\psfrag{b}{$6b$}
\psfrag{c}{$6c$}
\psfrag{d}{$6d$}
\begin{center}
{\scalebox{1}{
\includegraphics{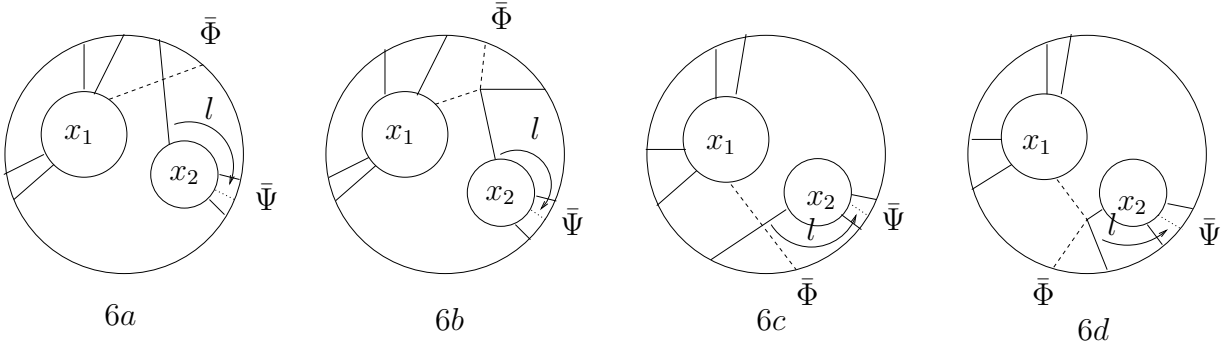}}
}
\end{center} 
\caption{Interacting diagrams for $\tilde{G}_3$ contributing to
$P_6$. Diagrams 6a and 6c have positive signs. Diagrams 6b and 6d have
negative signs.
}
\end{figure}

We now consider our second example of  three-point function
$\tilde{G_3}$. Considerations similar to the above lead us to the
following expression
\begin{equation}
\tilde{G_3} = \frac{1}{N \sqrt{J}} [1+ \frac{\d}{2} (\c +1 - \log 4
\pi -s) ]\cdot
\D(x_1)^{J_1+1}  \D(x_2)^{J_2+1}
(P_5 + \l K(x_1,x_2) P_6) .
\end{equation}
The only nontrivial difference from the previous case is encoded in
the factors $P_5$ and $P_6$.  
The factor $P_5$ is the sum of phase factors from the free diagrams
(see figure 5):
\begin{equation}
P_5 = \sum_{k=0}^{J_1} \sum_{l=0}^{J_2}
 e^{\frac{2 \pi i(k+l) n}{J}} 
= -\frac{J^2}{n^2 \pi^2} \sin^2(\frac{n \pi J_1}{J}).
\end{equation}
$P_6$ is the contribution of the phase factors from the interacting
diagrams shown in figure 6,
\bea
P_6 &=& 2 \sum_{l=0}^{J_2-1} 
e^{\frac{2 \pi i n l}{J}} (1- e^{\frac{2 \pi i n}{J}} )  
+ 2 \sum_{l=0}^{J_1-1} e^{\frac{2 \pi i n l}{J}} (1- e^{\frac{2 \pi i
n}{J}})\nn\\
&=& -\frac{J^2}{n^2 \pi^2} \sin^2(\frac{n \pi  J_1}{J}) 
\cdot \frac{-8 \pi^2 n^2}{J^2}.
\eea
We remark that the classes of diagrams similar to figure 3 and figure
4 cancel each other identically for the same reason as before.
 
Finally we obtain 
\begin{equation}
\tilde{G_3}(x_1,x_2) = 
\frac{\tilde{C}_{123} }
{
(4\pi^2 x_1^2)^{J_1+ 1 + \frac{\d}{2}}
(4\pi^2 x_2^2)^{J_2+ 1+ \frac{\d}{2}}
(4\pi^2 x_{12}^2)^{-\frac{\d}{2}}
},
\end{equation}
where the three-point function coefficient is
\begin{equation}
\tilde{C}_{123}  =  -\frac{J^{3/2}}{N}
\frac{1}{n^2 \pi^2} \sin^2(\frac{n \pi  J_1}{J}) \;
(1- \frac{\l' n^2}{2}).
\end{equation}

\section{Comparison with string theory}

Last week, 
using the formalism of \cite{GS,sv}, 
Huang \cite{huang} calculated the ratio on the LHS of \eq{hh} in the
large $\mu p^+ \a'$ limit and found agreement with the free field theory
expression. In this section, we will generalize his computation of the
string amplitudes to all orders in $1/(\mu p^+ \a')^2$ and
compare these new results with the field theory 
expressions derived in the last
section. We find perfect agreement to $O(\l')$. 

In the pp-wave background, the cubic string vertex is given by 
$| V \rangle = E_a E_b \ket{0}$, where $E_a$ and $E_b$ are the bosonic
and fermionic operators. For our purposes, we need only the expression
for $E_a$ \cite{sv}, 
\begin{equation}
E_a \sim \exp \left[ \frac{1}{2} \sum_{r,s =1}^3 
\sum_{m,n=-\infty}^\infty
a_{m(r)}^\dag \overline{N}^{(rs)}_{mn} a_{n(s)}^\dag
\right],
\end{equation}
where 
\begin{equation} \label{Neu}
\overline{N}^{(rs)}_{mn} = \d^{rs} \d_{mn} - 
2 \sqrt{\omega_{m(r)} \omega_{n(s)} } (X^{(r) {\rm T}} \Gamma_a^{-1}
X^{(s)})_{mn}
\end{equation}
are the Neumann matrices;
\begin{equation}  
\omega_{n(r)}=\sqrt{n^2+(\mu \alpha_{(r)})^2}, \quad \a_{(r)} := \a' p^+_{(r)}
\end{equation} 
are the oscillation frequencies of the $r$-th string, and the matrix
$\Gamma_a$ is given by
\begin{equation} 
\label{Ga}
(\Gamma_a)_{mn} = \sum_{r=1}^3 \sum_{p=-\infty}^\infty
\omega_{p(r)} X^{(r)}_{mp} X^{(r)}_{np}.
\end{equation}
The matrices $X^{(r)}$ arise from the  overlapping integrals in string
field theory, they were computed in \cite{sv} and are independent of
$\mu$. Thus the whole $\mu$ dependence is concentrated in the
frequencies $\omega_{m(r)}$. 

To the first order in $1/(\mu \a' p^+)^2 = \l'$, we have
\begin{equation}
\o_{n(3)} = \mu \a' p^+ (1+ \frac{\l' n^2}{2}),
\end{equation}
\begin{equation}
\o_{n(1)} = \k \mu \a' p^+ (1+ \frac{\l' n^2}{2\k^2}),
\end{equation}
\begin{equation}
\o_{n(2)} = (1-\k) \mu \a' p^+ (1+ \frac{\l' n^2}{2(1-\k)^2}),
\end{equation}
where $p^+ = p^+_{(3)}$. 
Here $\k$ is the ratio of the lightcone momentum of string 1 and 3,  
\begin{equation}
\k \equiv \k_1 := \frac{p^+_{(1)}}{p^+} = \frac{J_1}{J}, 
\quad \k_2:= \frac{J_2}{J} = 1- \k
\end{equation}
Our next goal is to compute the matrix $\Gamma_a$.
The leading order expression in the large $\mu \a' p^+$ limit 
was computed in \cite{huang} and reads
\begin{equation}
(\Gamma_a^{(0)})_{mn} = 2 \mu \a' p^+ \d_{mn}.
\end{equation}
At the next order in $\l'$, we find 
\begin{equation}
(\Gamma_a)_{mn} = (\Gamma_a^{(0)})_{mn}  + \frac{\l'}{2} 
\sum_l \left[\frac{l^2}{\k_1^2} \o_{l (1)}^{(0)} X^{(1)}_{ml}  X^{(1)}_{nl} +
(1 \leftrightarrow 2) \right] + \frac{\l'n^2}{4}
(\Gamma_a^{(0)})_{mn},
\end{equation}
where $\o_{l (r)}^{(0)} = \k_r \mu \a' p^+$ are the frequencies in the
lowest order in $\l'$. Using the explicit expressions for
$X^{(r)}_{ml}$ and summation formulae from the appendix D of
\cite{GS},  we obtain
\begin{equation} \label{final1}
(\Gamma_a)_{mn} = (\Gamma_a^{(0)})_{mn} + \mu \a' p^+ \l' n^2 \d_{mn}
=  (\Gamma_a^{(0)})_{mn} \; \left(1+ \frac{\l' n^2}{2}\right) . 
\end{equation}
Using this result, we derive the expression for the following 
elements of the  Neumann matrices,  
\begin{equation} \label{final2}
\overline{N}^{(3r)}_{n 0} = 
[\overline{N}^{(3r)}_{n 0}]^{(0)} \; \left(1 - \frac{\l'
n^2}{4}\right), \qquad r =1,2, 
\end{equation}
which will be needed for the analysis below.
The novelty in our results 
\eq{final1} and \eq{final2} lies in the second terms on the
RHS, which are the $O(\l')$ corrections to the zeroth order results of 
\cite{huang} and
\begin{equation} 
[\overline{N}^{(3r)}_{mn}]^{(0)}
= -\sqrt{\k_r} (X^{(3)T}X^{(r)})_{mn}.
\end{equation}

Now we consider the string scattering amplitudes which  correspond 
to the field theory three-point functions $G_3$ and $\tilde{G_3}$.
We have on the field theory side
\begin{equation} \label{53}
\frac{ C_{123} }{ C^{(\rm vac)}_{123} }  
= \frac{J}{J_1} 
\frac{1}{n^2 \pi^2} \sin^2(\frac{n \pi J_1}{J}) \cdot (1- \frac{\l' n^2}{2} ),
\end{equation}
\begin{equation} \label{54}
\frac{ \tilde{C}_{123} }{ C^{(\rm vac)}_{123} }  
= - \frac{J}{\sqrt{J_1 J_2}} \; 
\frac{1}{n^2 \pi^2} \sin^2(\frac{n \pi J_1}{J}) \cdot (1- \frac{\l' n^2}{2} ),
\end{equation}
where $C^{(\rm vac)}_{123} = \sqrt{J J_1 J_2}/N$. 

On the string side, we have for the first process (corresponding to \eq{53}),
\bea
\frac{ \langle \Phi_1| \langle \Phi_2|\langle \Phi_3| V \rangle }
{\langle 0_1| \langle 0_2|\langle 0_3| V\rangle } 
&=& \frac{1}{4} (\overline{N}^{(31)}_{n0} -\overline{N}^{(31)}_{-n0})^2 \\ 
&=& \frac{1}{4} ([\overline{N}^{(31)}_{n0} -\overline{N}^{(31)}_{-n0}]^{(0)})^2
\cdot (1- \frac{\l' n^2}{2} ) \nn \\
&=&  \frac{J}{J_1} 
\frac{1}{n^2 \pi^2} \sin^2(\frac{n \pi J_1}{J}) \cdot (1- \frac{\l' n^2}{2})
\eea
which is in perfect agreement with \eq{53}. Similarly for the second
string scattering (corresponding to \eq{54}), we obtain (cf. \cite{huang})
\bea
\frac{ \langle \Phi_1| \langle \Phi_2|\langle \Phi_3| V \rangle }
{\langle 0_1| \langle 0_2|\langle 0_3| V\rangle } 
&=& \frac{1}{2}  \overline{N}^{(31)}_{n0} \overline{N}^{(32)}_{n0} \\ 
&=& \frac{1}{2}  [\overline{N}^{(31)}_{n0} \overline{N}^{(32)}_{n0}]^{(0)}
\cdot (1- \frac{\l' n^2}{2}) \nn \\
&=&  - \frac{J}{\sqrt{J_1 J_2}} 
\sin^2(\frac{n \pi J_1}{J}) \cdot (1- \frac{\l' n^2}{2} )
\eea
which is again in agreement with our field theory result. 

Finally we generalize our string computations to all orders in
$\l'$. To do this,  we note that using the formulae in the appendix,
it is easy to derive
\begin{equation} \label{final3}
(\Gamma_a)_{mn} =  
2 \mu \a' p^+ \sqrt{1+ \l' n^2 } \d_{mn} = 2 \o_{n (3)} \d_{mn}. 
\end{equation}
This was obtained by substituting the all-orders expansion of the
square root in $\omega_{n(r)}=\sqrt{n^2+(\mu \alpha_{(r)})^2}$ in the
large $\mu$ limit into \eq{Ga}.  The resulting expression involves an
infinite sum arising from the multiplication of matrices of infinite
dimension. In order to work with
well-defined expressions, the sum has to be regularized. As
standard in string theory, we use the zeta function
regularization, as outlined in the Appendix.   
It then follows from \eq{Neu} that
\begin{equation} \label{final4}
\overline{N}^{(3r)}_{nm}= [\overline{N}^{(3r)}_{nm}]^{(0)} \; 
\left( \frac{1+ \l' m^2/\k_r^2}{1+ \l' n^2} \right)^{1/4}, \quad r= 1, 2.
\end{equation}
The results \eq{final3}, \eq{final4} generalize \eq{final1},
\eq{final2} to all orders in $\l'$. 
Equation \eq{final3} was derived by
resumming all-orders expansions in the large $\mu$ (small $\l'$)
limit. These expansions  contained only integer powers of $\lambda'$.
Our analysis is complete (in the sense that it does not miss any terms)
in the vicinity of $\lambda'=0$ and it cannot be analytically extrapolated 
to the opposite regime of $\lambda'=\infty$. 
 Even though our result \eq{final3} has a
closed analytic form, it is interpreted as an asymptotic expansion\footnote{ 
After this work appeared, it was argued in \cite{SV2} that terms
proportional to $(\l')^{3/2}$ will typically appear on the string
side, which is puzzling from the field theory point of view. We do not see 
such terms in our analysis.
}
in powers of $\l'$. 
This allows us to compute the corresponding three-string amplitudes
to all orders in small $\l'$. 
Note that the direct use of \eq{final3} 
to the opposite regime of large $\l'$ is not allowed, and in fact 
the Neumann matrices for $\l' = \infty$ are known and cannot be
obtained from  our formulae. 

Finally, this leads to the field theory predictions
\begin{equation} \label{61}
\frac{ C_{123} }{ C^{(\rm vac)}_{123} }
= \frac{J}{J_1}
\frac{1}{n^2 \pi^2} \sin^2(\frac{n \pi J_1}{J}) \cdot (1 + \l'n^2 )^{-1/2},
\end{equation}
\begin{equation} \label{62}
\frac{ \tilde{C}_{123} }{ C^{(\rm vac)}_{123} }
= - \frac{J}{\sqrt{J_1 J_2}} \;
\frac{1}{n^2 \pi^2} \sin^2(\frac{n \pi J_1}{J}) \cdot (1 + \l'n^2)^{-1/2}.
\end{equation}
It would be interesting to verify this
all-orders  prediction 
from the field theory point of
view.

\section*{Appendix: Summation formulae}

We note the useful identity \cite{CJ}
\begin{equation}
\sum_{l= -\infty}^\infty (-1)^l \frac{e^{i l y}}{l+ \b} = \frac{\pi}{ \sin
(\b\pi)} e^{- i \b y}, 
\quad -\pi < y < \pi,
\end{equation}
from which one can derive
\begin{equation}
\sum_{l=1}^\infty \frac{(-1)^l}{l^2-\b^2} \cos(l y)= \frac{1}{2\b^2} -
\frac{\pi}{2 \b \sin (\b \pi )} \cos (\b y), 
\end{equation}
and formally
\begin{equation} \label{l2p1}
\sum_{l=1}^\infty \frac{l^{2p}}{l^2 -\b^2} = -\frac{\pi}{2} \b^{2p-1}
\cot(\pi \b), \quad p \geq 1,
\end{equation}
\begin{equation} \label{l2p2}
\sum_{l=1}^\infty \frac{l^{2p}}{(l^2 -\b^2) (l^2 -\c^2)} = 
-\frac{\pi}{2 (\b^2 -\c^2)} (\b^{2p-1} \cot(\b \pi) - \c^{2p-1}
\cot(\c \pi)), \quad p \geq 1,
\end{equation}
\begin{equation} \label{l2p3}
\sum_{l=1}^\infty \frac{l^{2p}}{(l^2 -\b^2)^2} = \frac{\pi}{4}
[\pi \b^{2p-2} \csc^2(\b \pi) - (2p-1) \b^{2p-3} \cot(\b \pi)], \quad p \geq 1.
\end{equation}
The formulae \eq{l2p1}, \eq{l2p2} and \eq{l2p3} are 
understood by an appropriate analytic continuation using zeta function
regularization. In more details, define
\begin{equation}
f(s) := \sum_{l=1} ^\infty \frac{1}{l^s}\frac{1}{l^2 -\b^2} 
= \sum_{k=0} ^\infty \beta^{2k} \zeta(s+2+2k).
\end{equation}
Then we obtain \eq{l2p1}
\begin{equation}
f(-2p) 
= -\frac{\pi}{2} \b^{2p-1} \cot(\pi \b). 
\end{equation}
Equations \eq{l2p2} and \eq{l2p3} follow immediately.

\section*{Acknowledgements} 
We would like to thank Simon Ross, Rodolfo Russo and Adrian Signer for
useful discussions. We acknowledge grants from the Nuffield foundation
and PPARC.

\end{document}